\documentclass[aps,pra,superscriptaddress,amsfonts,amsmath,amssymb,reprint,showpacs,floatfix]{revtex4-1}

\usepackage{url}
\usepackage{bm}
\usepackage{graphicx}
\usepackage{amsmath}
\usepackage{amstext}
\usepackage{amssymb}
\usepackage{amsfonts}
\usepackage{amsbsy}
\usepackage{verbatim}
\usepackage{color}
\usepackage[colorlinks=true, urlcolor=blue, linkcolor=blue, citecolor=blue, pdftex]{hyperref}
\usepackage{multirow}
\usepackage[final]{pdfpages}
\usepackage{floatrow}
\newcommand\beq            {\begin{equation}}
\newcommand\bea           {\begin{equation}\begin{array}l\displaystyle}
\newcommand\eeq            {\end{equation}}
\newcommand\bes           {\begin{subequations}}
\newcommand\esu           {\end{subequations}}

\newcommand{\ees}{\end{split}}
\newcommand{\eea}{\end{eqnarray}}

\begin{document}

\title{Quantum-statistics-induced flow patterns in driven ideal Fermi gases}

\author{Marco Beria}
\email[]{marco.beria@sissa.it}
\affiliation{SISSA$-$International School for Advanced Studies and INFN, Sezione di Trieste, Via Bonomea 265, I-34136 Trieste, Italy, EU}

\author{Yasir Iqbal}
\email[]{yiqbal@ictp.it}
\affiliation{The Abdus Salam International Centre for Theoretical Physics, P.O. Box 586, I-34151 Trieste, Italy, EU}

\author{Massimiliano Di Ventra}
\email[]{diventra@physics.ucsd.edu}
\affiliation{Department of Physics, University of California San Diego, La Jolla, California 92093, USA}

\author{Markus M\"uller}
\email[]{markusm@ictp.it}
\affiliation{The Abdus Salam International Centre for Theoretical Physics, P.O. Box 586, I-34151 Trieste, Italy, EU}

\date{\today}

\begin{abstract}
\noindent
While classical or quantum interacting liquids become turbulent under sufficiently strong driving, it is not obvious what flow pattern an {\it ideal} quantum gas develops under similar conditions. Unlike classical
noninteracting particles which exhibit rather trivial flow, ideal fermions have to satisfy the exclusion principle, which
acts as a form of {\it collective} repulsion.
We thus study the flow of an ideal Fermi gas as it is driven out of a narrow orifice of width comparable to the Fermi wavelength, employing
both a microcanonical approach to transport, and solving a Lindblad equation for Markovian driving leads. Both methods are in good agreement and predict an outflowing current density with a complex microscopic pattern of vorticity in the steady state. Applying a bias of the order of the chemical potential results in a short-range correlated antiferromagnetic vorticity pattern, corresponding to local moments of the order of a tenth of a magneton, $e\hbar/2m$, if the fermions are charged. The latter may be detectable by magnetosensitive spectroscopy in strongly driven cold gases (atoms) or clean electronic nanostructures (electrons).
\end{abstract}

\pacs{03.75.Ss, 05.60.Gg, 05.30.Fk, 67.10.Hk}

\maketitle

\section{Introduction}

Experimental advances in the rapid quenching and imaging of ultracold atoms~\cite{Greiner-2001,Kohl-2005,Bloch-2008,Polkovnikov-2011,Brantut-2012,Stadler-2012}, as well as in the microscopy of nanoscale structures~\cite{Topinka-2000}, make it possible to observe interesting effects of current-carrying systems with specific features attributable to their quantum nature. On the other hand, it has long been known$-$ever since the beginning of quantum mechanics~\cite{Madelung-1926}$-$that interacting or noninteracting quantum fluids can be described in terms of hydrodynamic equations~\cite{Bloch-1932, Martin-1959, Vignale-1997, Tokatly-2005,DAgosta-2006}. In fact, the many-body
time-dependent Schr\"odinger equation is {\it exactly} equivalent to the equations of motion for
the density and velocity field, namely the continuity equation and a hydrodynamic equation$-$albeit with an unknown stress tensor~\cite{Madelung-1926,Martin-1959,mybook}. As any viscous fluid, interacting quantum liquids can exhibit a transition between laminar and turbulent flows when driven strongly~\cite{DAgosta-2006,Bushong-2007a}, and experiments on electron liquids to detect this phenomenon have been proposed~\cite{Bushong-2007b, Muller-2009, Dyakonov-2005}.
It has been argued that electronic systems close to quantum criticality are best suited to exhibit such turbulence, as they feature a small ratio of shear viscosity and entropy density, undoped graphene~\cite{Muller-2009,Mendoza-2011}
or cold atoms at unitarity above $T_c$~\cite{Enss} being probably the simplest such systems.

More specific effects in the flow of {\em quantum} fluids have been studied, both experimentally and theoretically~\cite{Paoletti-2011,Tsubota-2009,Nemirovskii-2013,Khalatnikov,Pitaevskii-1961,Gros-1963,landau}, in the context of vorticity and turbulence in superfluids, focusing on the dynamics and the interactions of quantized vortices.
However, another interesting aspect of {\em fermionic} quantum fluids has not received much attention so far. Complex flow patterns are generally attributed to the particles' interactions which generate nonlinearity and chaoticity in the dynamics. With the advent of atomic gases where interactions among atoms can be tuned to essentially zero~\cite{Bloch-2008}, the
question arises whether the exclusion principle alone$-$a form of {\it collective} repulsion$-$may lead to interesting flow patterns, if the Fermi gas is driven
out of equilibrium. This is in contrast to ideal {\em classical} gases, which definitely do not have any interactions to produce a finite viscosity and the associated turbulent
effects~\cite{landau}. However, that nontrivial phenomena do arise in out-of-equilibrium flow of free fermions was recently shown in 1$d$ gases~\cite{Hunyadi-2004,Bettelheim-2006,Bettelheim-2012,Protopopov-2013}, where the formation of shocks and interference ripples after a local quench was observed, suggesting that even more interesting phenomena could develop in higher dimensions. Given that already in equilibrium, Pauli exclusion leads to interesting interference phenomena in free fermions, such as Friedel oscillations~\cite{Friedel-1952}, one may expect even more complex patterns out of equilibrium.

In this paper we study the simplest case exemplifying these phenomena: driven ideal fermions in restricted geometries in {\it two} dimensions. We show that they indeed develop nontrivial vorticity patterns, which are manifestations of the Fermi statistics. Antiferromagnetic patterns are found not only in transients (unlike in the above-mentioned $1d$ studies), but also in the long-time steady state. The latter should facilitate their observation in experiments, such as most recent transport measurements in cold atoms where constrictions as we consider here have been realized~\cite{Brantut-2012,Stadler-2012}. Note instead that the ideal Bose gas, if prepared in a condensed state at $T=0$, reduces to a single-particle problem, so that particular effects of ``anti-exclusion" are absent. However, we recall that the case of free bosons is a pathological limit, which is likely to be modified by any weak interaction, as it alters the low-energy spectrum and thus ensures superfluidity, by establishing a finite critical velocity.\\

The paper is organized as follows. In Sec.~\ref{sec:2} we introduce the model and the methods to be employed. In Sec.~\ref{sec:3} we present the main physical results, while their experimental verification is discussed in Sec.~\ref{sec:4}. Conclusions are given in Sec.~\ref{sec:5}.

\begin{figure}
\includegraphics[width=1.0\columnwidth]{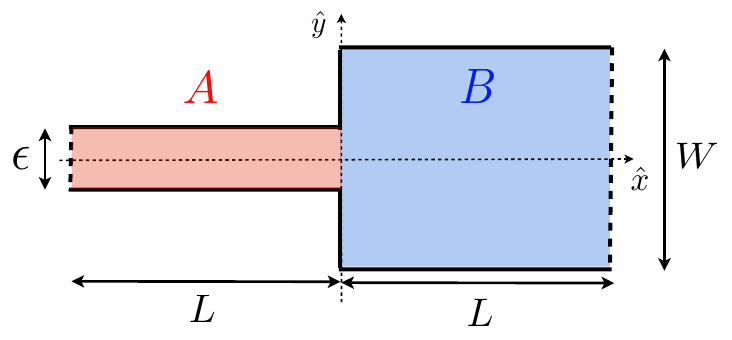}
\caption{(Color online) Studied setup: a narrow channel ({\it A}) connected to a wide region ({\it B}). In the microcanonical setup the dashed boundaries at $\pm L$ are hard walls, while they represent Markovian leads in the Lindblad approach. For large $L$, the steady-state pattern near the orifice may be expected to be the same in the two approaches.}
\label{fig:geometry}
\end{figure}

\begin{figure*}[t]
\includegraphics[width=1.0\columnwidth]{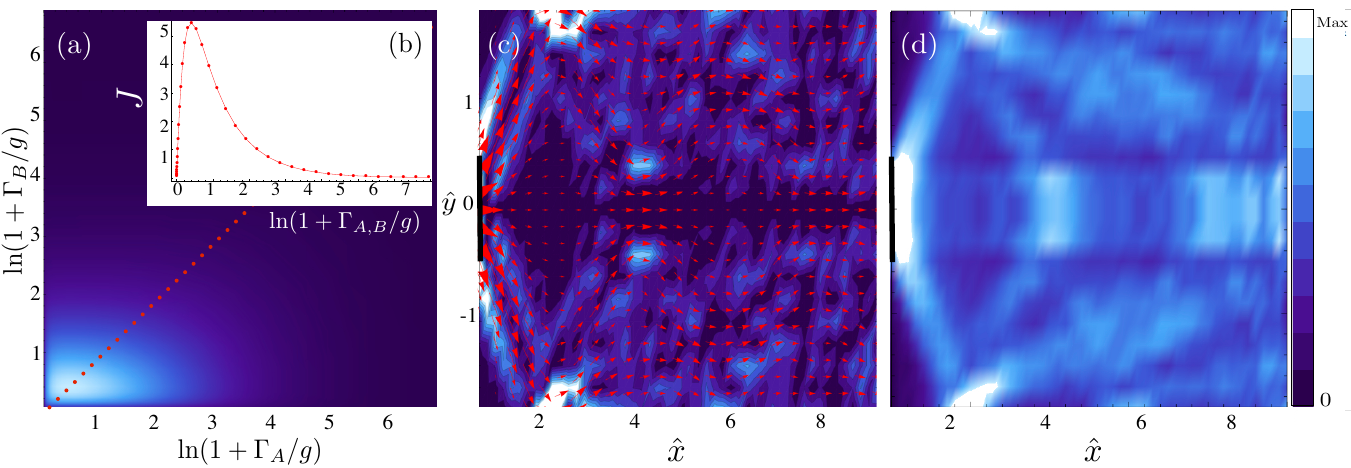}
\caption{(Color online)  Steady state from the Lindblad equation at high average density, $\bar{n}\epsilon^{2}\approx 2.5$. (a) Total current $J$ in units of $\frac{\hbar}{2m}\bar{n}^{3/2}\epsilon$ as a function of the bath couplings $\Gamma_{A,B}/g$. (b) Current as a function of driving, evaluated along the dashed red line ($\Gamma_{A}=\Gamma_{B}$). At  $\Gamma\gtrsim g$, the dependence is nonmonotonic. (c) Current density field $\textbf{J}(x,y)$ (red arrows) superimposed on a contour plot of $|\nabla\times {\textbf J}|$, for $\Gamma_{A,B}=0.256 g$.
The orifice is marked by a solid bold line. (d) Corresponding density pattern. Distances are measured in units of the orifice width $\epsilon$ [but note that the aspect ratio is $2$:$5$ in (c) and (d)]. In all plots lighter colors correspond to larger values (linear scale).}
\label{fig:linear}
\end{figure*}

\section{Model}
\label{sec:2}
In this work, we consider a two-dimensional Fermi gas in the setup shown in Fig.~\ref{fig:geometry}: a narrow channel ($A$) of width $\epsilon$ is connected at $x=0$ to a much wider region ($B$) of width $W=4\epsilon$ (in all figures, if not specified otherwise). We study the current flow, as the out-of-equilibrium free Fermi gas is driven from $A$ to $B$. We are interested in the limit of very long regions, $L\rightarrow\infty$. This problem is solved using two complementary approaches: the microcanonical formalism~\cite{DiVentra-2004,DiVentra-2012} for a closed system, and the Lindblad equation in what was called the ``third quantization'' formalism~\cite{Prosen-2008,Prosen-2010} for a driven open system. While the first method allows us in principle to follow the entire dynamics at every instant of time, the second one accesses directly the steady-state properties.

\subsection{Microcanonical approach}
We prepare fermions of mass $m$ in the ground state of the Hamiltonian $H_0=-\frac{\hbar^2}{2m}{\bf \nabla}^2 + V_{\rm in}\theta(x)$ at chemical potential $\mu$, in the setup of Fig.~\ref{fig:geometry}, where $\theta(x)$ is the Heaviside step function. This results in average densities $\bar n_{A}$ and $\bar n_{B}$ in regions $A$, and $B$, respectively. We consider  $L \gg1/\bar n_{A}\epsilon$, so as to simulate a quasi-infinite system. At $t=0$, we suddenly quench the potential to $H=-\frac{\hbar^2}{2m}{\bf \nabla}^2 + V_{\rm fin}\theta(-x)$, and study the ensuing time evolution. As explained below, we only work with combinations of biases $(V_{\rm in},~V_{\rm fin})$ for which either the initial or the final potential is zero. In all numerical implementations we discretize the system on a square lattice with lattice constant $a$, using a tight-binding model with hopping $g=\frac{\hbar^2}{2ma^2}$. \\

In order to highlight the effect of quantum statistics within the microcanonical approach, we compare two different protocols.
 The first protocol, which we refer to as {\it expansion into fermions}, starts from an initial state with a roughly equal average density throughout the closed geometry. This is obtained by fixing the initial potential to $V_{\rm in}=0$ which establishes a chemical potential $\mu$ in the sample. At $t=0$, a bias of the order of the chemical potential, $V_{\rm fin}\approx \mu/2$, is turned on in region $A$, so as to push the fermions out of the narrow channel into the large box $B$. Under the influence of this bias, the fermions are forced to exit from the narrow channel, and flow into the Fermi sea, which is already present in the box $B$. Note that the density profile of the latter is not entirely uniform, but oscillates in the transverse direction, because of the boundary conditions imposed at $y=\pm W$.

In the second protocol, which we refer to as {\it expansion into free space}, the initial state is prepared by applying a large potential $V_{\rm in}\gg \mu$ to region $B$, such that  the gas is initially confined to the narrow channel only. At $t=0$, the potential is released, and the gas is left to expand freely ($V_{\rm fin}=0$) into the empty region $B$.

We will compare these two protocols to highlight the role of finite density, and its oscillations in region $B$. The latter are relevant only in the {\it expansion into fermions} protocol. The quasisteady states obtained within that protocol are the closest to the steady states that one can realize within driven open systems, which will be addressed in the next subsection.

We are interested in a quasisteady current flow in the vicinity of the orifice, within a large time window $\hbar/\mu \lesssim t \lesssim L/v_F$, where $v_F \approx \sqrt{2\mu/m}$. In order to ensure this, we always choose the conserved number of fermions $N$, and the potentials $V_{\rm in/fin}$ such that the narrow channel is populated with a finite density in the initial state. In an {\it expansion into fermions} where no potential is applied in the initial state, i.e., $V_{\rm in}=0$, a finite initial density in channel $A$ requires that $\bar n_B \epsilon^2>\frac{\pi}{4}$ (assuming that $\bar n_B W^2\gg 1$, which we always ensured). Since the wave function remains a Slater determinant at all times, it is enough to solve for the time-dependent single-particle eigenvalues, and eigenfunctions before the quench, $H_0\psi^{(0)}_\alpha(x,y)=E_\alpha\psi^{(0)}_\alpha(x,y)$, and afterwards, $H\psi_\beta(x,y)=E_\beta\psi_\beta(x,y)$. To this effect, both the pre- and postquench Hamiltonians $H_0$ and $H$, respectively, are {\it exactly} diagonalized, and the time-dependent density and current density are obtained as
\begin{eqnarray}
\langle\hat n(x,y,t)\rangle &=&\sum_{\alpha\in {\rm occ}}\sum_{\beta,\beta'}\psi^{*}_{\beta}(x,y,t)\psi_{\beta'}(x,y,t)\Theta_{\beta\alpha}^{*}\Theta_{\beta'\alpha},\,\nonumber\\
\langle\hat{ \textbf{J}}(x,y,t)\rangle =&\frac{\hbar}{m}&{\rm Im}\Bigg[\sum_{\alpha\in {\rm occ}}\sum_{\beta,\beta'}\psi^{*}_{\beta}(x,y,t) \Bigg.\nonumber \\
\quad\quad\quad\quad \Bigg. &\times&{\bf \nabla}\psi_{\beta'}(x,y,t)\Theta_{\beta\alpha}^{*}\Theta_{\beta'\alpha}\Bigg],
\label{currentdensity}
\end{eqnarray}

\noindent where
\begin{eqnarray}
\Theta_{\alpha\beta}=\langle \psi^{(0)}_\alpha | \psi_\beta\rangle
\end{eqnarray}
are overlaps between eigenstates of the initial and final Hamiltonian, and the time evolution is simply given by
\begin{eqnarray}
\psi_\beta(x,y,t)=\psi_\beta(x,y)e^{-iE_\beta t}.
\end{eqnarray}
The summation over $\alpha$ in Eq.~(\ref{currentdensity}) is restricted to the set of states (labeled as ${\rm occ}$) that are occupied in the initial Fermi sea. Herein lies the central difference with a system of ideal bosons, where only the lowest energy state would be occupied, which leads to rather trivial patterns. We found numerically that at any fixed position close to the orifice, after an initial transient a steady state establishes, a result which is in fact not obvious for free fermions. \\

\subsection{Lindblad equation} In order to directly access the nonequilibrium steady-state properties and compare with the microcanonical results we consider again the setup of Fig.~\ref{fig:geometry}, but now taking  the dashed boundaries to represent couplings to driving Markovian baths. The density matrix $\rho(t)$ of this {\it open} quantum system evolves according to a master equation, expressed in Lindblad form as~\cite{Prosen-2008,Prosen-2010}
\begin{equation}
\label{eqn:lindbladeqn}
\frac{d\rho}{dt}=\hat{\mathcal{L}}\rho:=-\frac{i}{\hbar} [H_0,\rho] + \sum_{b,y}(2L^b_{y}\rho L_{y}^{b \dagger}-\{L_{y}^{b \dagger}L^b_{y},\rho\}).
\end{equation}

\noindent
For convenience we directly formulate a discrete version of the problem on a lattice with spacing $a$ and hopping amplitude $g=\frac{\hbar^{2}}{2ma^{2}}$. The operators $L^b_{y}$, which are linear in the fermions, represent the coupling at coordinate $y$ on boundary $b \in \{A,B\}$ to independent baths. This allows for an exact solution via the method of third quantization~\cite{Prosen-2008,Prosen-2010}. To describe driving from $A$ to $B$ we choose $L^{A}_y=\sqrt{\Gamma_{A}}c^{\dagger}~(x=-L,y)$ and $L^{B}_y=\sqrt{\Gamma_{B}}c~(x=L,y)$, where $c^\dagger,c$ are creation and annihilation operators, respectively.

\section{Results}
\label{sec:3}
 In this section we present physical results obtained within the two formalisms, and compare them.
Our main result is the existence of two density regimes, connected by a smooth crossover in $\bar n \epsilon^2$. In the  {\em high-density} regime ($\bar n \epsilon^2\gg1$) the Fermi wavelength is small as compared to the channel width $\epsilon$, and the flow behavior can be explained by semiclassical diffraction theory. In contrast, in the {\em low-density} regime $\bar n\epsilon^2\sim 1$, the Fermi wavelength is comparable with the channel width and the flow exhibits peculiar patterns that result from interference and effects of Pauli exclusion, i.e., the fermionic statistics.

\begin{figure*}[t]
\includegraphics[width=1.0\columnwidth]{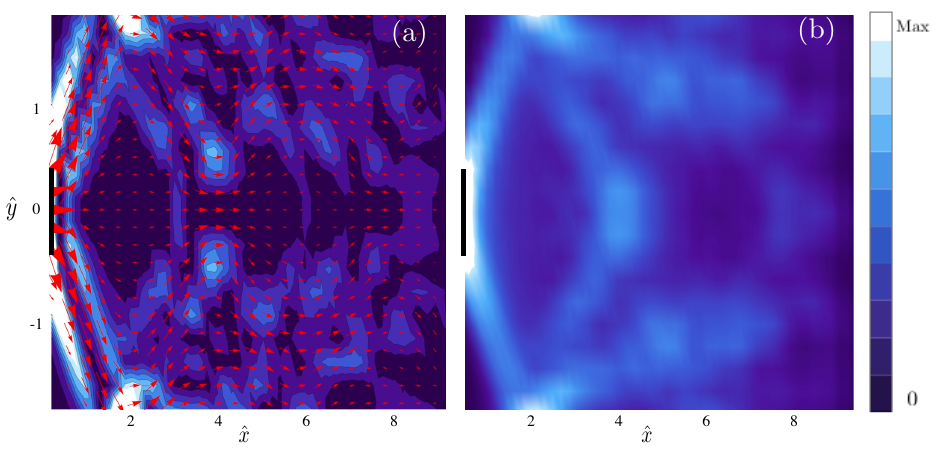}
\caption{(Color online) Quasisteady state in the high-density regime ($\bar n_A\epsilon^2\approx 5$), evaluated within the microcanonical formalism (free expansion into an initially empty region $B$). The observed patterns, shown for the late time $t\approx250\hbar/\mu$, are qualitatively very close to the steady state found in open systems.
(a) Current density superimposed on the absolute value of the vorticity, and (b) density. $x$ and $y$ coordinates are in measured in units of the orifice width $\epsilon$. Note the aspect ratio $2$:$5$ of the axis scales.}
\label{fig:micro-linear}
\end{figure*}

\subsection{Quasisteady state in the microcanonical approach}
\begin{figure*}
\includegraphics[width=1.0\columnwidth]{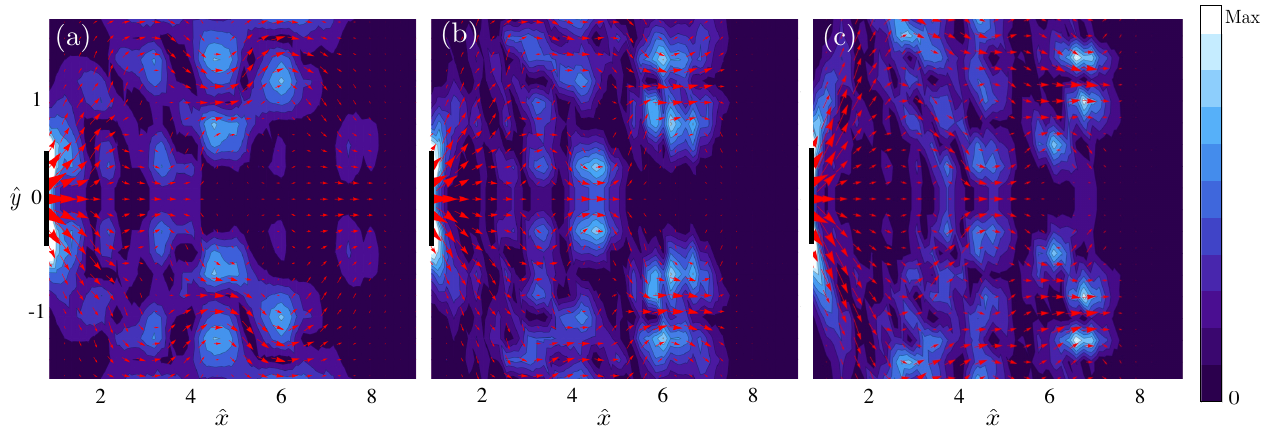}
\caption{(Color online) Current density $\textbf J$ and contour plot of $|\nabla \times {\textbf J}|$ computed in the microcanonical quasisteady state, for increasing density. From left to right: $\bar{n}\epsilon^{2}\approx1.1$, $\bar{n}\epsilon^{2}\approx 2.2$, $\bar{n}\epsilon^{2}\approx 4.8$. The most interesting patterns are observed at relatively low density, $\bar{n}\epsilon^{2}\sim 1$.}\label{fig:non-trivial}
\end{figure*}

Within the microcanonical formalism, we found that, after a short transient time following the quench, the formed current and vorticity patterns remain remarkably stable under time evolution. Only much later, when waves reflect from the boundaries at $x=\pm L$ flow back to the junction, this pattern is obviously disturbed [cf. Ref.~\cite{supplementary} for a visualization of the initial dynamics which result eventually in the nontrivial quasisteady flow pattern of Fig.~\ref{fig:non-trivial}(a)]. This observation strongly suggests that a genuine {\em quasisteady state} forms within our microcanonical setup, which is well defined and infinitely long lived in the thermodynamic limit $L\to \infty$. Note that this is a nontrivial result, especially for noninteracting particles, since it is not obvious that in the absence of interactions and randomization of momenta a steady state should establish. The existence of such a steady state makes it meaningful to ask for the equivalence between this microcanonical setup and the transport formalism based on a driven open system.

\subsection{Equivalence of transport approaches}
The steady state of the open system and the quasisteady state that establishes in the microcanonical approach (both within the {\it expansion into fermions} and {\it expansion into free space} protocols) are usually closely related, except close to the boundaries. In fact, the two approaches to transport are expected to be equivalent in the thermodynamic limit, if interactions and impurity scattering induce thermalization and momentum randomization far from the junction. However, for noninteracting fermions the equivalence of the two methods is not guaranteed, since reflected waves from the contacts at $x=\pm L$ can coherently propagate back to the junction. Despite this caveat we found that the two approaches give quantitatively very similar results in the high-density regime $\bar{n}\epsilon^{2}\gg1$, where reflections appear to be of minor importance, whereas in the low-density regime the agreement is only qualitative. In that regime we expect the microcanonical approach to be a better description for a realistic system with weak but finite interactions. We therefore present our central results for the low-density regime within that framework.

\subsection{High-density regime}
We first discuss the high-density regime, which features rather simple semiclassical diffraction patterns. At high densities, $\bar{n}\epsilon^{2}\gg1$, the Fermi wavelength is much smaller than the width of the orifice. Hence, a semiclassical diffraction picture is expected to hold. Accordingly, we find a relatively simple steady-state pattern. For a representative set of parameters $\Gamma_A/g=\Gamma_B/g=0.256$, the current density pattern and the absolute value of its vorticity ($|\nabla \times \textbf{J}|$) are shown in Fig.~\ref{fig:linear}(c), as well as the density pattern in Fig.~\ref{fig:linear}(d). In Fig.~\ref{fig:linear}(a), we show the total current $J$
in the plane of couplings to the leads. Interestingly, $J$ is nonmonotonic in $\Gamma_{A,B}$. Along every line of constant ratio $\Gamma_A/\Gamma_B$ it reaches a maximum and decreases as $1/\Gamma_{A,B}$ at strong coupling [see Fig.~\ref{fig:linear}(b)], similarly as was observed in related studies in $1d$~\cite{Prosen-2009,Benenti-2009,Zunkovic-2010,Karevski-2009,Prosen-notes,negdiffbook}.

The most prominent feature of the current pattern in Fig.~\ref{fig:linear}(c) is a diffraction beam exiting from the orifice, whose angle is determined by the transverse momentum of the highest propagating band in channel $A$ (here the second band). Those beams are reflected at $y=\pm W/2$ and give rise to two islands of intense vorticity around $(4\epsilon,\pm\epsilon/2)$. They are thus simple boundary effects due to reflections at finite $W$; see also the density pattern in Fig.~\ref{fig:linear}(d). The main characteristics of this regime is a rhomboidal region of very small vorticity centered at $(2\epsilon,0)$. The microcanonical approach reproduces all these features at high densities, $\bar{n}\epsilon^{2}\gg1$.
In particular, for both protocols we defined in the microcanonical ensemble, we found matching conditions of the driven systems, such that density and current-density patterns in the respective steady states coincided.

\begin{figure}[h]
\includegraphics[width=1.0\columnwidth]{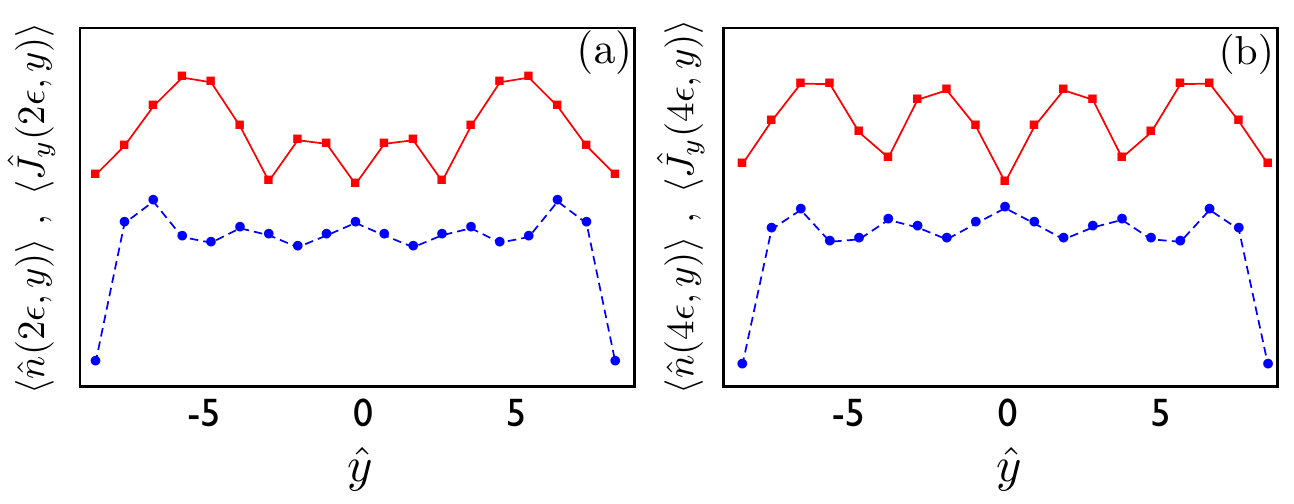}
\caption{(Color online) The profile of the spatial density (dashed blue line) and the $y$ component of the current density (solid red line) along the cross sections at $x=2\epsilon$ (a), and $4\epsilon$ (b). Density oscillations and the current are anticorrelated. The corresponding values have been rescaled to visually pronounce the anticorrelated behavior.}
\label{fig:j_y}
\end{figure}

In Fig.~\ref{fig:micro-linear} we show the result for an {\it expansion into free space} (for reasons of illustration only), with a finite initial density $\bar{n}_{A}\epsilon^2\approx5$ in channel $A$. The gas is then left to expand freely (with ${\rm V}_{\rm fin}=0$) into the empty region $B$. The most important characteristics of the vorticity pattern is again the empty rhomboidal region of size $\sim W^2$ close to the orifice. This is a main feature of the semiclassical high-density  regime, where only simple  diffraction at the orifice is observed. It results in two beams, which, however, produce little interesting interference and vorticity patterns.
Nevertheless, the very good agreement between the microcanonical and the Lindblad approach to transport is an interesting result in itself. It suggests that the two approaches can be used essentially interchangeably to describe steady states out of equilibrium, up to the caveat that in a fully coherent system, interference with reflections from the far boundaries may result in differences, as we will discuss below in the low-density regime. The latter seems to be of negligible importance in the high-density regime, however. Note that this equivalence of approaches is quite analogous to the equivalence of thermodynamic ensembles in equilibrium statistical mechanics. However, while such an equivalence appears rather natural in the context of interacting, fully chaotic systems, it is much less obvious for free fermions, as we discussed above. For the latter even the existence of a quasisteady state in the closed system could not be anticipated {\it a priori}~\cite{mybook,DiVentra-2004}. It would be interesting to understand this phenomenon from the perspective of quantum chaos in many-fermion systems.

\begin{figure*}
\includegraphics[width=1.0\columnwidth]{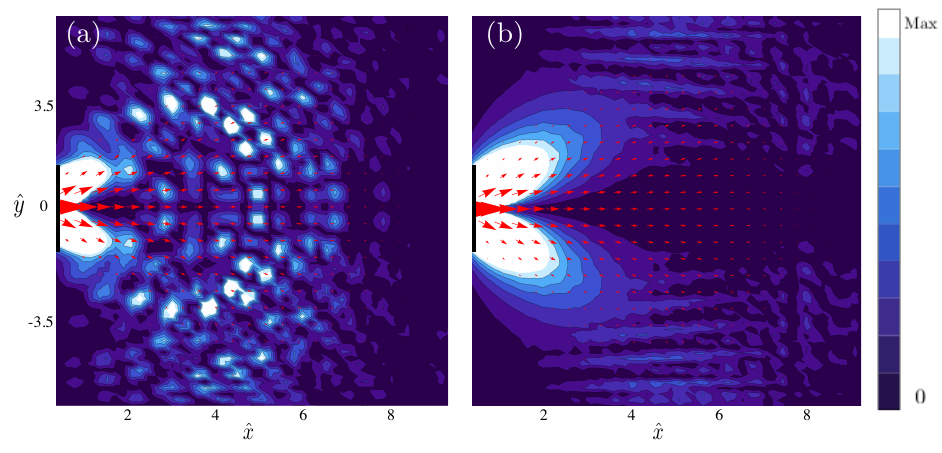}
\caption{(Color online) (a) The current-density pattern and the intensity of the corresponding vorticities are shown for $\bar n \epsilon^2\approx 1.3$ in the microcanonical quench protocol (snapshots are taken at time $t\approx16\frac{\hbar}{\mu} $): (a) $V_{\rm in}=0$ (roughly homogeneous initial density), $V_{\rm fin}\approx\mu/2$ applied to region $A$; (b) expansion into empty space, $V_{\rm in}\gg1$ (region $B$ initially empty), $V_{\rm fin}=0$. The vorticity patterns in the two cases are evidently very different. In particular, in case (b) of the expansion into the initially empty region $B$, there is no evidence of nontrivial features in the current flow and vorticity close to the orifice, in contrast to case (a). This shows that a finite density in region $B$ and the associated Friedel oscillations in the steady state are important for the formation of nontrivial flow patterns. In both cases, we chose $W=14\epsilon$ so that the boundaries at $\pm W$ are rather far from the orifice and the difference between the two quench protocols is evident. The color code is the same for both figures. Note the aspect ratio of $7$:$5$ of the axis scales.}
\label{fig:f_nf}
\end{figure*}

\subsection{Low-density regime}

The flow pattern is much more interesting at low density, $\bar{n}\epsilon^{2}\sim1$, where quantum effects are more pronounced. In this regime, additional structures in the current and vorticity develop within the rhomboidal region, which was nearly vorticity free at high density. In Fig.~\ref{fig:non-trivial}(a), we show the current and vorticity pattern computed within the microcanonical formalism at time $t\approx 16\hbar /\mu$, before reflections at $x=\pm L$ occur. Here, we study the protocol corresponding to {\it expansion into fermions}. The emerging steady-state current pattern exhibits preferential zigzag-shaped stream lines, and thus differs markedly from the simple picture predicted by semiclassical diffraction theory. As one should expect, the spatial scale of the vorticity variations is set by the interparticle distance $\sim\frac{1}{\sqrt{\bar{n}}}$, i.e., the Fermi wavelength. Note that here the islands of vorticity close to the orifice ($x<4\epsilon$) are {\it not} simply due to the reflection of outflowing waves from the boundaries at $y=\pm W/2$.

We interpret the origin of the complex current patterns as arising from current flow through regions that exhibit boundary induced Friedel oscillations in the steady-state density. These oscillations are provoked by a finite lateral confinement (finite $W$), even though $W^2\bar{n}$ may be fairly large. The current flow appears to avoid regions of higher density in the steady state, which leads to nontrivial vorticity patterns. This is illustrated in Fig.~\ref{fig:j_y}, which shows that in an expansion into fermions, the $y$ component of the current density is strongly anticorrelated with the density oscillations in region $B$ in the steady state. This suggests that  one may view both effects as consequences of Pauli exclusion, which leads to Friedel oscillations in the density {\em and} the currents, in the steady state.

The patterns farther from the orifice ($x\gtrsim 4\epsilon$) depend, naturally, rather strongly on the presence of the boundaries, as they are dominated by interference of waves that are reflected from the boundaries. If instead one considers periodic boundary conditions in the $y$ direction, the patterns are dominated by interference of waves that wind around the cylinder. However, near the orifice, the effect of changing boundary conditions is much less pronounced. Even though the vorticity pattern is modified quantitatively, it remains qualitatively similar (see Appendix~\ref{sec:bc}).

Even though from Eq.~(\ref{currentdensity}) it is clear that the current density is simply the superposition of single-particle contributions, the resulting pattern is quite nontrivial, as forward and backward propagating states superpose, each with their individual interference patterns.
The higher the initial density the larger is the number of superposed modes, which tends to smoothen the interference patterns. Interesting patterns survive for low-density $\bar n\epsilon^2\gtrsim \pi/4$, while in the limit $\bar n\epsilon^2\to \infty$ we expect to recover  the classical limit where all nontrivial features are smeared out. We illustrate this trend in Fig.~\ref{fig:non-trivial}, where the current patterns obtained for increasing densities are shown. Nontrivial structures emerge at low density where the Fermi wavelength is comparable with the orifice width, provided that current is flowing into a nonempty region $B$. Under these circumstances the Pauli exclusion leads to boundary induced Friedel oscillations in region $B$, which appear to play a crucial role, and induce the observed vorticity patterns.

The importance of an appreciable density of fermions in region $B$ and the associated  density oscillations in the steady state can be best appreciated by comparing an {\it expansion into fermions} with an {\it expansion into free space}. We work at low density ($\bar n \epsilon^2\approx 1.3$) within the microcanonical formalism. Figure~\ref{fig:f_nf}(a) shows the steady-state pattern of current and vorticity in an {\it expansion into fermions}. 

We contrast the above protocol with an expansion into empty space, with very similar initial density in region $A$. Figure~\ref{fig:f_nf}(b) shows the resulting flow pattern, which exhibits hardly any interesting features. We interpret this as being due to the low steady-state density of fermions in region $B$, such that density oscillations in that area are very weak and have little effect on the current flow. The main conclusion from this comparison is that it is not merely the geometry that is relevant for producing interesting interference patterns in driven fermions, but also the presence of an appreciable steady-state density in the relevant spatial regions.

For a driven open system, we obtain similar steady-state properties, but the details of the flow patterns differ, due to the pronounced role of reflections from the boundaries at $x=\pm L$, which act as a semitransparent wall causing partial reflection of the particle flux. These reflected waves are also the cause of minor differences in the density patterns of Figs.~\ref{fig:linear}(d) and Fig.~\ref{fig:micro-linear}(b). Consequently, the reflected and incoming waves interfere to form a complicated structure of currents and vorticities (see Appendix~\ref{sec:3qld} for the corresponding figures and discussion).

\section{Possible experimental verification}
\label{sec:4}
If the fermions are charged, such as in a $2d$ electron gas (which may still be considered weakly interacting in the presence of a strong dielectric) a complex current pattern as in Fig.~\ref{fig:non-trivial}(a) generates a nontrivial magnetic-field distribution. To obtain the same, one defines

\begin{eqnarray}
\label{BiotSavart}
{\bf b}({\bf r}=\{x,y,z\}) = \frac{\mu_0}{4\pi} \int dx'dy'  {\bf j}({\bf r}') \times \frac{{\bf r}-{\bf r}'}{|{\bf r}-{\bf r}'|^3},
\end{eqnarray}

\noindent where $\mu_0$ is the permeability of vacuum, and ${\bf j}$ is the $2d$ number current density of atoms in the $x$-$y$ plane ($z=0$). It then follows from the Biot-Savart law that the magnetic field ${\bf B}({\bf r})$ generated by moving particles of charge $e$ is given by

\begin{eqnarray}
{\bf B}({\bf r}) = e {\bf b}({\bf r}).
\end{eqnarray}

\begin{figure}
\includegraphics[width=1.0\columnwidth]{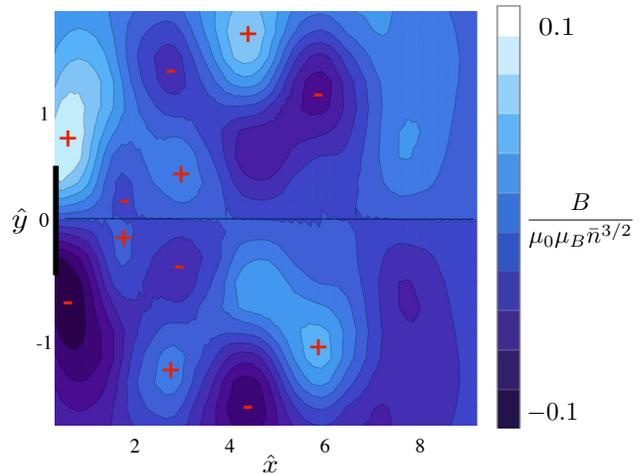}
\caption{(Color online) The $z$ component of the magnetic field generated by a charged current flow as in Fig.~\ref{fig:non-trivial}(a): The quasisteady state exhibits a {\it staggered} flux close to the orifice.}
\label{fig:B}
\end{figure}

For strong drivings ($V_{\rm fin}\sim \mu$) in the considered geometry, typical magnetic moments associated with the circulation patterns are of the order of a tenth of a magneton $\mu=\frac{e\hbar}{2m}$, that is, in principle, an experimentally accessible intensity. Interestingly, the vorticity maxima organize  in a short-range correlated antiferromagnetic pattern, which realizes an out-of-equilibrium {\it staggered} flux state, cf. Fig.~\ref{fig:B}, reminiscent of equilibrium {\it staggered} flux phases proposed in strongly correlated $2d$ systems~\cite{Affleck-1988,Marston-1989}. Similar patterns arise from currents of magnetically (electrically) polarized neutral atoms. The electric (magnetic) fields due to such moving dipoles are proportional to a derivative of the field pattern of Fig.~\ref{fig:B}, and are given by

\begin{eqnarray}
{\bf E}({\bf r}) = -({\bf m}\cdot {\bf \nabla}) {\bf b}({\bf r}).
\end{eqnarray}

\noindent with ${\bf E}({\bf r})$ being the electric field, and ${\bf m}$ the static magnetic moment. Similarly, the magnetic field produced due to polarized neutral particles with a static electric dipole ${\bf d}$ is given by

\begin{eqnarray}
{\bf B}({\bf r}) = ({\bf d}\cdot {\bf\nabla}) {\bf b}({\bf r}).
\end{eqnarray}

\begin{figure*}
\includegraphics[width=1.0\columnwidth]{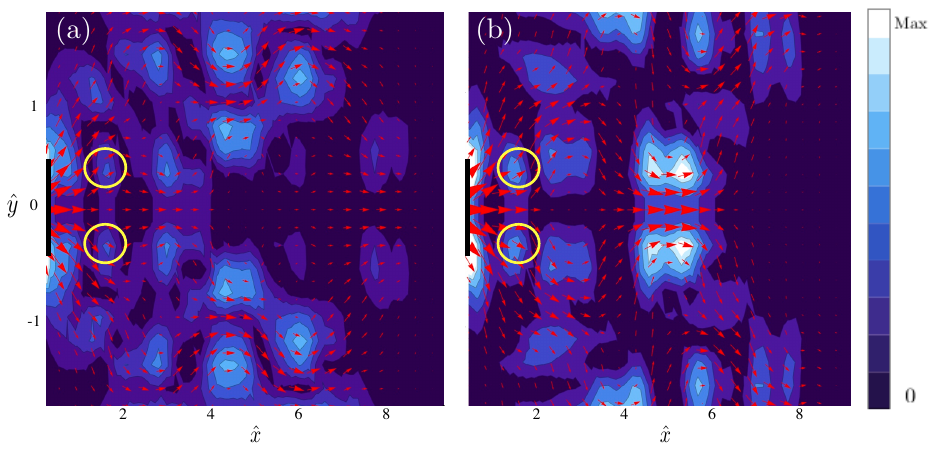}
\caption{(Color online) Comparison between (a) closed and (b) periodic boundary conditions at $y=\pm W/2$ (here $W=4\epsilon$) in the low-density regime ($\bar n\epsilon^2\approx1.1$). Shown is the current density, superimposed on the vorticity contour plot, as obtained within the microcanonical approach, with $V_{\rm in}=0$, and $V_{\rm fin}\approx\mu/2$.
Two nontrivial patterns close to the orifice at $(x,y)\approx(2\epsilon,\pm \epsilon/2)$ are encircled in yellow.
The $x$ and $y$ coordinates are measured in units of the orifice width $\epsilon$. Note the aspect ratio $2$:$5$ of the axis scales. The color code is the same for both figures.}
\label{fig:bc}
\end{figure*}

\noindent However, these fields may be too weak to be detected by present experimental means. It may be interesting to look for similar patterns in systems which have spin-orbit coupling, e.g., in cold atoms.

Apart from the currents, the density patterns computed in this paper can be measured experimentally by resonant light absorption in atomic gases in optical lattices. The limiting resolution is currently~~$\sim 660$ nm~\cite{Esslinger}, which is smaller than typical Fermi wavelengths in those systems. The setup discussed here has already been realized in recent experiments~\cite{Brantut-2012,Stadler-2012} where $^6$Li atoms are confined in a geometry with a narrow constriction  connecting two reservoirs.

\begin{figure}
\includegraphics[width=1.0\columnwidth]{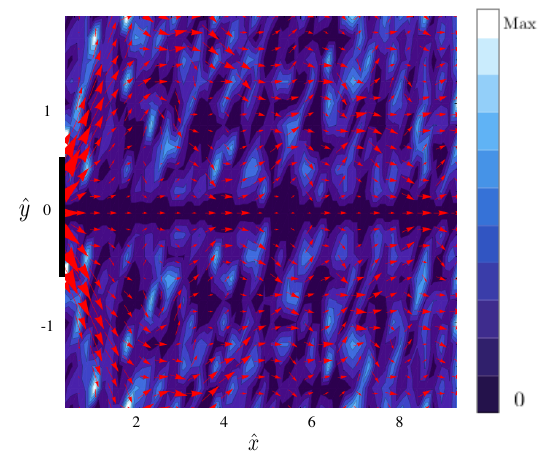}
\caption{(Color online) The current density superimposed on the absolute value of the vorticity, as obtained with a low average density, $\bar n\epsilon^2\approx 1.1$. In this regime quantum interference effects are strong, as in closed systems, but they differ quantitatively, because of important reflections at the leads $x=\pm L$, which are absent in the quasisteady state analyzed in closed systems.
$x$ and $y$ coordinates are in measured in units of the orifice width $\epsilon$. Note the aspect ratio $2$:$5$ of the axis scales.}
\label{fig:non-linear-3q}
\end{figure}

\section{Conclusion}
\label{sec:5}
Using two complementary approaches, we have analyzed both the steady-state properties and the transient dynamics of an ideal Fermi gas pushed out of an orifice into a wider region. Pauli exclusion was found to strongly influence the current flow of fermions at finite density: it induces current patterns with staggered local moments of appreciable size, formed by itinerant fermions in an out-of-equilibrium steady state. The latter may be used to experimentally probe the predicted patterns. Since these effects are after all interference phenomena, we expect them to be robust towards weak interactions. It would be interesting to extend the present study to disordered systems and compare with the predictions of very heterogeneous current flow with substantial steady vorticity therein~\cite{Aronov-1986}. We expect that density inhomogeneities due to Friedel oscillations from strong impurities (which take the role of the walls at $\pm W$) will lead to similar interesting vorticity patterns under a nonequilibrium steady state.
\\
\\
{\em Acknowledgments} M.D. acknowledges support from DOE Grant No. DE-FG02-05ER46204. We thank S. Kehrein and T. Prosen for useful discussions.

\appendix

\section{INFLUENCE OF BOUNDARY CONDITIONS}
\label{sec:bc}
We analyzed the influence of boundary conditions on the formation of nontrivial current patterns in closed systems, by considering different boundary conditions. We used the microcanonical formalism with $V_{\rm in}=0$, and $V_{\rm fin}\approx\frac{1}{2}\mu$, in the low-density regime $\bar n \equiv \bar n_B \approx \bar n_A \approx 1.1/\epsilon^{2}$, where nontrivial patterns appear.
In Fig.~\ref{fig:bc}(b) we show patterns obtained with periodic boundary conditions where $y=\pm W/2$ are identified. We observe that the main {\it qualitative} features of the patterns close to the orifice ($x \lesssim  4 \epsilon\sim W$) are still present with periodic boundary conditions; in particular, the two islands of vorticity close to $(x,y)\approx(2\epsilon,\pm\epsilon/2)$ still form. However, one should not expect quantitative agreement near the orifice.

\section{STEADY STATE OF OPEN SYSTEMS IN THE LOW-DENSITY REGIME}
\label{sec:3qld}
In an open system the current flow is controlled by the strength of the couplings $\Gamma_{A,B}$ to the external leads. The low-density regime, where the Fermi wavelength is of the order of the orifice width, is obtained, e.g., by tuning the injection rate $\Gamma_{A}$, such that $\frac{\Gamma_{A}}{g}\ll 1$.

For this regime, the current and vorticity patterns of the steady state are shown in Fig.~\ref{fig:non-linear-3q}. The rhomboidal region of size $\sim W^2$ close to the orifice contains several intense local maxima of the vorticity, in qualitative agreement with what is found using the microcanonical approach to closed systems.  However, along with this feature a much more complicated structure of vorticity develops throughout the wide region $B$. This arises because the coupling to the absorbing bath at $x=L=10\epsilon$ acts only as a semitransparent wall (it becomes fully transparent only in the limit of infinite absorption $\Gamma_B\rightarrow\infty$). These effects due to reflected waves survive even close to the junction, because of the lack of dephasing and randomization in this noninteracting system. While some reflection is certainly also present in the higher density regime, its relative effect is apparently much smaller, so that the driven open system and the quasisteady state of the closed system are qualitatively very similar.

In contrast to the involved structure of the steady state of the open system, the microcanonical approach can access the quasisteady current pattern {\em before}  reflections from the boundary at $\pm L$ occur. It is thus more suitable to reveal the effects induced on the current pattern by quantum statistics, and separating them from simple reflection effects. The most interesting effects due to Pauli exclusion are found at low densities $\bar n_A\epsilon^2\gtrsim 1$, in the rhomboidal region close to the orifice where the reflection of diffracted beams from either boundary do not play much of a role.

\end{document}